\documentclass[
    ,final            
  ]
  {aipproc}

\layoutstyle{6x9}

\begin{document}

\title{EXPLORE/OC: A Search for Planetary Transits in the Field of NGC 2660}

\author{Kaspar von Braun}{
  address={Department of Terrestrial Magnetism, Carnegie Institution of
Washington, \\5241 Broad Branch Road, Washington, DC 20015}
}

\author{Brian Lee}{
  address={Dept of Astronomy and Astrophysics, University of Toronto,
  \\60 St. George St., Toronto, Canada M5S 3H8}
}

\author{Gabriela Mall\'{e}n-Ornelas}{
  address={Harvard-Smithsonian Center for Astrophysics, \\60 Garden Street, 
  MS-15, Cambridge, MA 02138}
}

\author{Howard Yee}{
  address={Dept of Astronomy and Astrophysics, University of Toronto,
  \\60 St. George St., Toronto, Canada M5S 3H8}
}

\author{Sara Seager}{
  address={Department of Terrestrial Magnetism, Carnegie Institution of
Washington, \\5241 Broad Branch Road, Washington, DC 20015}
}

\author{Michael Gladders}{
  address={Carnegie Observatories, \\813 Santa Barbara St.,
  Pasadena, CA 91101}
}


\begin{abstract}
We present preliminary photometric results of a monitoring study of
the open cluster NGC 2660 as part of the EXPLORE/OC project to find
planetary transits in Galactic open clusters. Analyzing a total of
21000 stars (3000 stars with photometry to 1\% or better) yielded
three light curves with low-amplitude signals like those typically
expected for transiting hot Jupiters. Although their eclipses are most
likely caused by non-planetary companions, our methods and photometric
precision illustrate the potential to detect planetary transits around
stars in nearby open clusters.
\end{abstract}

\maketitle


\section{Introduction}

As part of the
EXPLORE\footnote{\url{http://www.ciw.edu/seager/EXPLORE/explore.htm}}
Project \cite{msy03}, we have recently begun a survey of southern open
clusters (OCs) with the aim of detecting planetary transits around
cluster member stars
(EXPLORE/OC\footnote{\url{http://www.ciw.edu/seager/EXPLORE/open_clusters_survey.htm}}). Probing
cluster populations provides a complement to our ongoing deep
monitoring studies of rich Galactic fields \cite{msy03}.

\noindent Open cluster monitoring provides the following advantages and incentives:
\begin{itemize}
\item In general, metallicity, age, distance, and foreground reddening are
either known or may be determined for cluster members (more easily
than for random field stars). Thus, planets detected around cluster
stars will readily represent data points for any statistic correlating
planet frequency with age or metallicity of the parent star.
\item The planet-formation processes, and hence planet frequencies, may
differ between the open cluster and Galactic field populations. This
study allows the EXPLORE Project to compare these two different
environments.
\item Specific masses and radii for cluster stars may be targeted in the
search by the choice of cluster and by adjusting exposure times for
the target. In general, smaller stars offer better chances to detect
the low-amplitude transit signal.
\end{itemize}

\noindent The difficulties and challenges involving open cluster surveys are:

\begin{itemize}
\item The number of monitored stars is typically lower than in rich Galactic
fields, reducing the statistical chance of detecting planets.
\item Determining cluster membership of stars in the open cluster fields
without spectroscopic data is difficult due to the contamination by
Galactic field stars. Since the clusters are typically concentrated
toward the Galactic disk, this contamination may be significant.
\item Significant differential reddening across the cluster field and
along the line of sight can make isochrone fitting (and subsequent
determination of physical parameters such as age, distance, and
metallicity) difficult.
\end{itemize}

Note that both of the latter two difficulties may at least in part be
circumvented by obtaining spectra. In this work, we illustrate some of
the points mentioned above, describe our observing and data-reduction
strategies, and show some of our preliminary results of the southern
open cluster NGC 2660. Preliminary results of the study of our second
target, NGC 6208, are presented in a companion paper \cite{lbm04} in
this volume.

\section{Data}

\begin{figure}
  \includegraphics[height=.3\textheight]{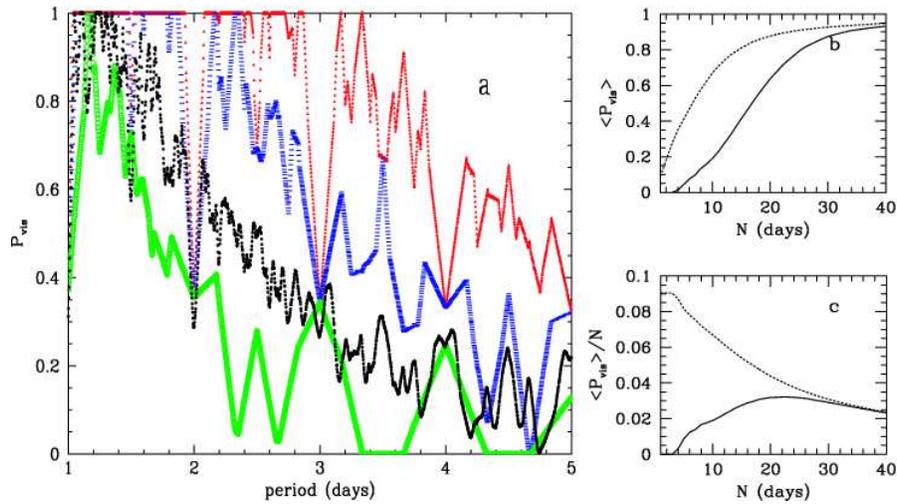}

  \caption{Probability $P_{vis}$ of detecting existing transiting
planets with different orbital periods. $P_{vis}$ is calculated with
the requirement that two transits must be observed.  {\bf Panel a}:
$P_{vis}$ of detecting 2 transits of an existing transiting planet
with a period between 2 and 5 days after 21 (top curve), 14 (second
curve from the top) and 7 (bottom curve) consecutive, uninterrupted
nights of observing (10.8 hours per night). The difficulty of
detecting some phase angles is shown by the dips in the curves (e.g.,
orbital periods of an integer number of days may always
feature their transits during the day and are therefore statistically
harder to detect). All phases are considered for each period. The
second curve from the bottom shows the real $P_{vis}$ for our
monitoring study of NGC 2660 (19 nights of 7-8 hours per night, with
interruptions due to weather and telescope scheduling; see Fig. 3).
{\bf Panel b:} The mean $P_{vis}$ as a function of number of
consecutive nights in an observing run (10.8 hour nights). The solid
line is for the requirement to detect two transits and the dashed line
for one transit. This figure indicates how much the likelihood
of finding existing transits grows with an increasing number of
nights of observing.  {\bf Panel c:} The efficiency of
<$P_{vis}$> per night. For the two-transit requirement (solid line)
and 10.8 hour long nights, an observing run of 21 nights is most
efficient. For the single transit requirement, the efficiency
decreases monotonically with the number of nights since additional
nights have progressively lower probabilities of detecting "new"
transits.}
\end{figure}

Our observations of NGC 2660 ($l = 265.9^{\circ}; b=-3.01^{\circ}$)
were obtained with the Carnegie Institution's Swope 1m Telescope at
the Las Campanas Observatory in Chile during the nights of 2003
Feb. 10--28 which were partially hampered by poor weather
conditions. Our field of view was approximately 23' $\times$ 15' in
size.  Only $I$-band data with a cadence of around 7 minutes were
obtained for monitoring. In that filter, a transit will be most easily
distinguishable from contaminants such as grazing binaries because
color-dependent limb-darkening results in steeper ingress and egress
as well as a flatter eclipse bottom in $I$ than in other bands
\cite{sm03}. We show in Fig. 1 our observing window function which
indicates the likelihood of our detecting {\it existing} transits and
measuring their periods to within an aliasing factor (we require at
least two transits). A necessary condition for detecting existing
transits is, of course, high-precision photometry \cite{msy03}. In
Fig. 2, we show our photometric precision as a function of magnitude.

\begin{figure}
  \includegraphics[height=.3\textheight]{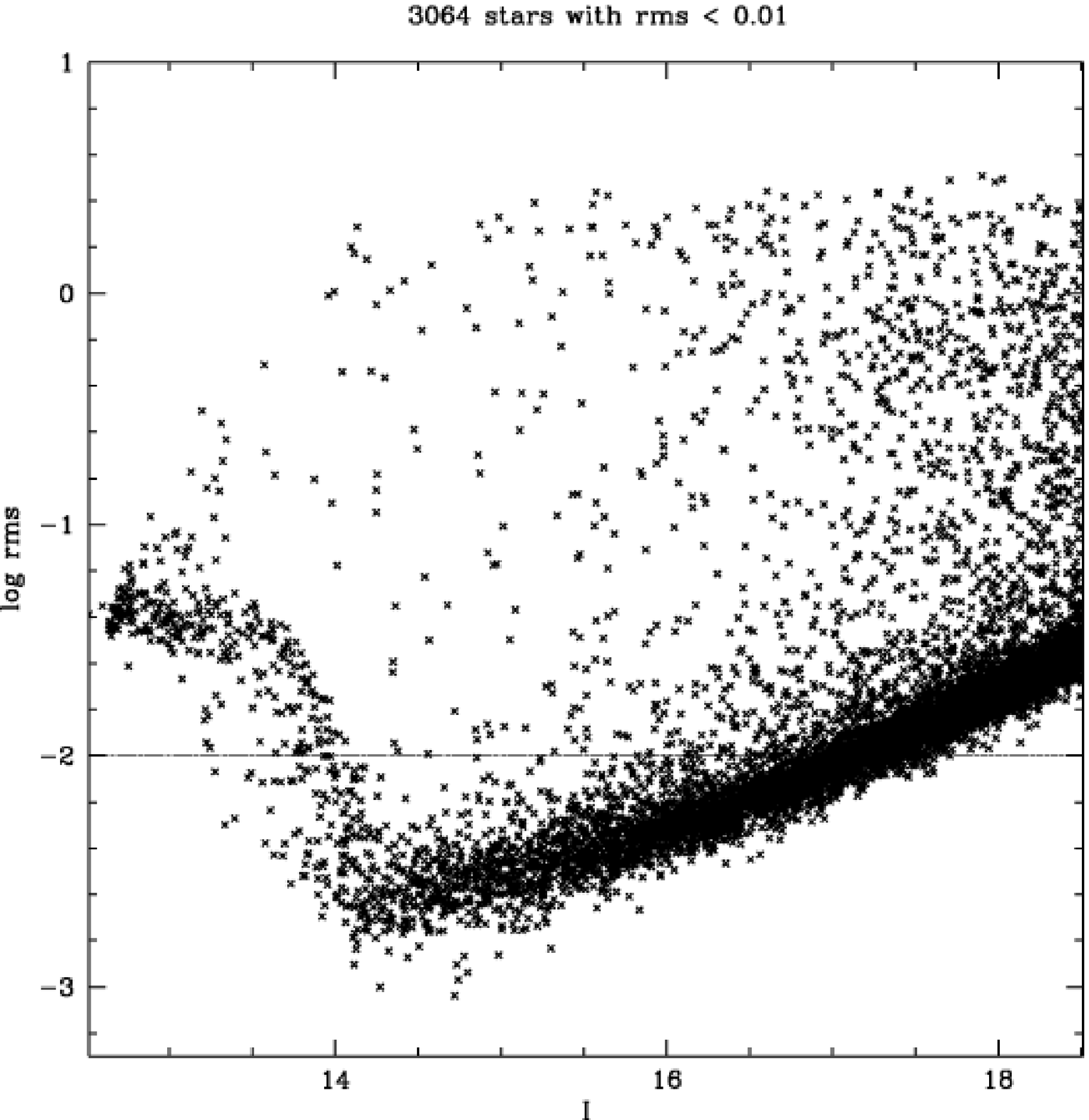}

  \caption{Photometric precision of night 1 of our monitoring run of
NGC 2660. Of the roughly 21000 monitored stars, slightly more than
3000 have photometry of precision 1\% or better. This rms is measured
as the scatter around the mean magnitude of the star under
investigation. The 1\%-photometry stars cover a magnitude range of
slightly more than 2.5 mags. By adjusting the exposure time, one can
therefore target OC member stars of a range of certain spectral types
to maximize the likelihood of detecting a transit (taking into account
distance to the cluster and foreground reddening).}
\end{figure}

\section{Results}

Around 7000 unphased photometric light curves of the stars with the
best photometric precision were visually inspected to detect
transit-like signals.  We show in Fig. 3 three examples of light
curves with the low-amplitude signature typically expected for
transiting hot Jupiters. Although the eclipses visible in the light
curves are most likely not due to planets, we are currently analyzing
spectral data on them to determine spectral types of the parent stars
as well as their variations in radial velocities. Regardless of
whether or not we are able to find a planetary transit in NGC 2660, we
have illustrated the potential of our methods to detect (unphased)
amplitude variations of less than 1\%, the most fundamental
requirement to detect planetary transits.

\begin{figure}
  \includegraphics[height=.4\textheight]{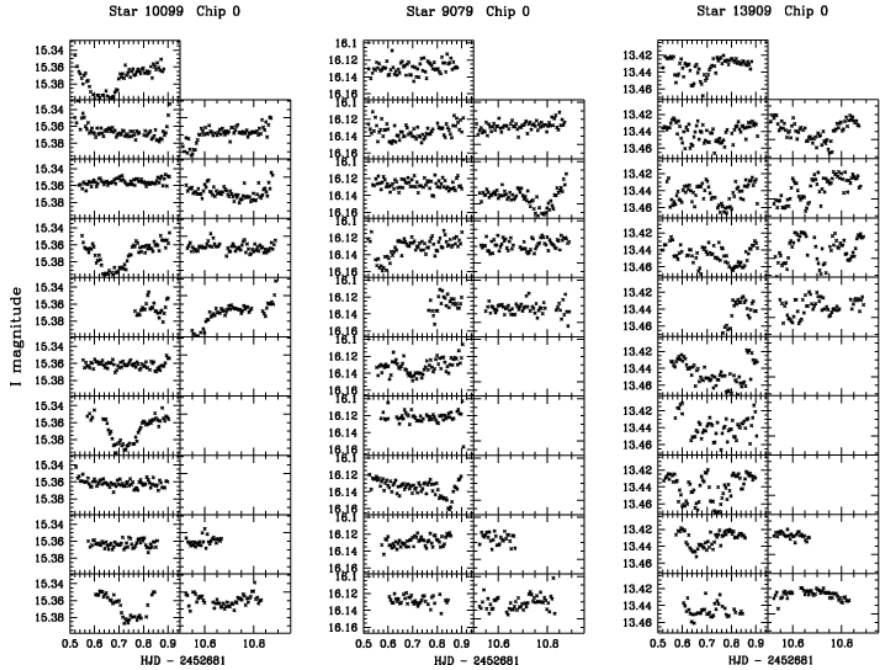}

  \caption{Examples of real-time (unphased) light curves from our
monitoring run of NGC 2660. Every panel represents data taken during
one night, starting on the bottom left with night 1. Night 2's data
are directly above it, night 3 above that and so on. We did not obtain
any data during nights 13-15. All three panels display the
low-amplitude signal that we are looking for in the search for
planetary transits even though they are most likely caused by a
larger-sized companion (left panel) or grazing binaries (middle and
right panels) \cite{sm03}.}
\end{figure}


\begin{theacknowledgments}
We would like to express our gratitude to the Las Campanas Staff for 
their unparalleled helpfulness and dedication to optimizing every little 
aspect of our observing runs.
\end{theacknowledgments}

\bibliographystyle{aipproc}

\end{document}